\documentclass[10pt,twocolumn,aps,prl,groupedaddress]{revtex4-1}

\usepackage{amsmath}
\usepackage{amssymb}
\usepackage{graphicx}
\usepackage{soul}

\def\bea{\begin{eqnarray}}
\def\eea{\end{eqnarray}}
\def\ben{\begin{equation}}
\def\een{\end{equation}}
\def\benu{\begin{enumerate}}
\def\enu{\end{enumerate}}

\def\lsim {\ifmmode {\buildrel<\over\sim}}

\def\sss{\scriptscriptstyle\rm}

\def\1var{(\bx_1...\bx\N)}

\def\half{\frac{1}{2}}

\def\br{{\bf r}}

\def\b1{{\bf 1}}
\def\bx{{x}}

\def\xc{_{\sss XC}}

\def\Hxc{_{\sss HXC}}

\def\N{_{\sss N}}
\def\H{_{\sss H}}

\def\sph_int{ {\int d^3 r}}

\def\infintd3r{ \int_{-\infty}^\infty d^3r\,}
\def\intd3r{ \int d^3r\,}

\def\laplace1d{\frac{d^2}{dx^2}}
\def\plaplace1d{\frac{d^2}{d{x'}^2}}

\def\padr2{\frac{\partial^2}{\partial r^2}}

\def\N{{\cal N}}

\def\b{{\beta}}

\begin{document}


\title{Density Functional Theory for Fractional Particle Number: Derivative Discontinuity of the Energy at the Maximum Number of Bound Electrons}


\author{Daniel L. Whitenack}
\email[]{dwhitena@purdue.edu}
\homepage[]{http://www.purdue.edu/dft}
\affiliation{Department of Physics, Purdue University, 525 Northwestern Avenue, West Lafayette, IN 47907, USA}
\author{Yu Zhang}
\email[]{yuz10@uci.edu (now at UC Irvine)}
\affiliation{Department of Chemistry, Purdue University, 560 Oval Drive, West Lafayette, IN 47907, USA}
\author{Adam Wasserman}
\email[]{awasser@purdue.edu}
\affiliation{Department of Chemistry, Purdue University, 560 Oval Drive, West Lafayette, IN 47907, USA}
\affiliation{Department of Physics, Purdue University, 525 Northwestern Avenue, West Lafayette, IN 47907, USA}


\date{\today}

\begin{abstract}
The derivative discontinuity in the exact exchange-correlation potential of ensemble Density Functional Theory (DFT) is investigated at the specific integer number that corresponds to the maximum number of bound electrons, $J_{max}$.  A recently developed complex-scaled analog of DFT is extended to fractional particle numbers and used to study ensembles of both bound and metastable states.  It is found that the exact exchange-correlation potential experiences discontinuous jumps at integer particle numbers including $J_{max}$.  For integers below $J_{max}$ the jump is purely real because of the real shift in the chemical potential.  At $J_{max}$, the jump has a non-zero imaginary component reflecting the finite lifetime of the $(J_{max}+1)$ state.  
\end{abstract}


\maketitle


\section{Introduction}
\begin{figure}
  \scalebox{0.5}{\includegraphics{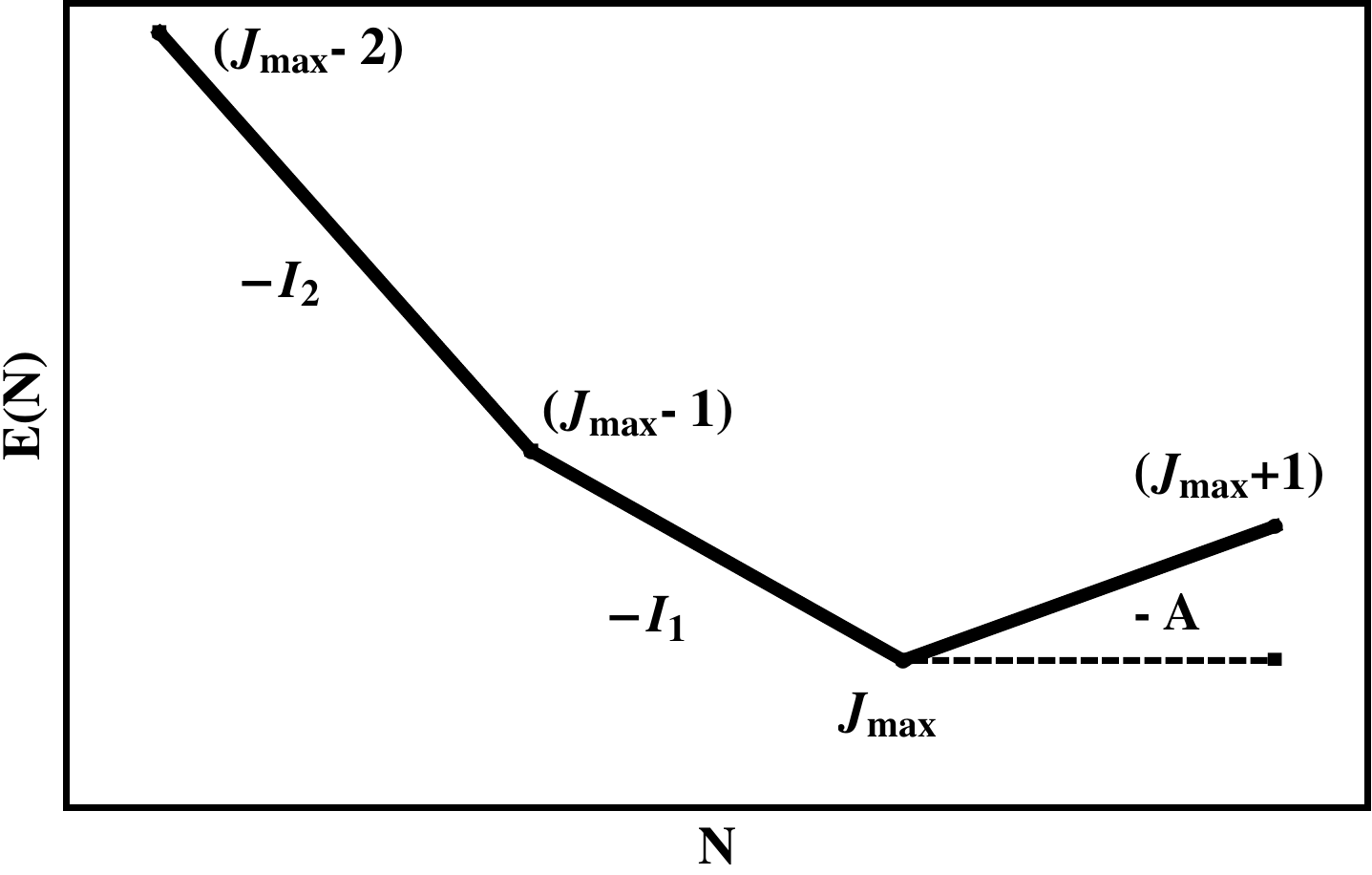}}
  \caption{Cartoon illustrating the ground-state energy of an open system as a function of particle number.  If a long-lived metastable state exists above the maximum number of electrons ($J_{max}$), the curve can be thought to turn upward because of the negative electron affinity of the $J_{max}$-electron system.}
  \label{fig:cartoon}
\end{figure}

The ground-state energy of an open system as a function of particle number has derivative discontinuities at integer values and is linear in between integers~\cite{PPLB82,PL83,SS83} (see Fig.~\ref{fig:cartoon}). Since Density Functional Theory (DFT)~\cite{hohenberg,kohn} is formally exact, an extension of the theory to fractional particle numbers should also include this behavior.  Most approximate functionals used in practice have non-linear behavior between integers and lack derivative discontinuities in the energy.  These incorrect features of approximate functionals have profound consequences \cite{CMY08}. The associated delocalization and static correlation errors typically lead to an underestimation of chemical barriers, band gaps, long-range charge transfer excitations, and energies of dissociating molecules and an overestimation of binding energies.

The derivative discontinuities in the energy have been studied at integer particle numbers lower than the maximum number of bound electrons. For example, by considering the 2-component ensemble formed by a $J$-electron system and the corresponding $(J+1)$-electron system in an exactly-solvable 2-electron model system, Sagvolden and Perdew established that the exact Kohn-Sham exchange-correlation potential experiences a constant positive jump at integer $J=1$ \cite{SP08}.   They also postulated that one could study the discontinuity at the integer number, $J_{max}$, of particles that corresponds to the maximum number of bound electrons.  However, their useful analysis could not be extended to $J_{max}$ because the $(J_{max}+1)$ electron system is not bound.  The discontinuity, on the other hand, is a property of the potential at the integer value where the system \emph{is} bound, so it would be useful to study such behavior.

The electron affinity of a $J$-electron system is defined as the energy difference, $A=E_{J} - E_{J+1}$, when one electron is added to the system.  In the case when $J=J_{max}$, the affinity could be taken as zero because the $(J_{max}+1)$ system is not stable (dashed line in Fig.\ref{fig:cartoon}).  However, the $(J_{max}+1)$ system could form a long-lived metastable anion, and such a system is often described as having a ``negative electron affinity'' (NEA)~\cite{S08,S11}.  Therefore if the energy is plotted as a function of particle number ($N$), below $J_{max}$ we have a concave up curve with discontinuities at the integer values ($J$'s) and negative slopes, but beyond $J_{max}$ the curve can be thought to turn upward because of the change in sign of the electron affinity (see Fig~\ref{fig:cartoon}).  One physical system that exhibits this behavior is the nitrogen molecule ($J_{max}=14$) whose negative ion is unbound but long-lived.  The NEA of N$_2$ is -2.3 eV or -0.084 Hartree because of the existence of a long-lived (14+1)-electron metastable state ($N_2^-$) 2.3 eV above the ground state of N$_2$.  The width of this $N_2^-$ state is 0.57 eV corresponding to a lifetime of 1.2 femtoseconds.

When a molecule with $J_{max}$ electrons is in contact with an infinite but distant metallic reservoir of work function $W$ smaller than the magnitude of the NEA of the molecule ($W<|A|$, e.g. N$_2$ in contact with a slab of Cs), the ground-state density of the combined system is particularly interesting. At infinite separation, it is simply the sum of neutral, isolated N$_2$ and metallic densities. At finite separation, however, close to the molecule, it is given by a linear combination of the $J_{max}$-electron ground-state density (for N$_2$) and the $(J_{max}+1)$-electron density of metastable N$_2^-$. 

As mentioned, the lack of the correct derivative discontinuities in approximate functionals has a dramatic impact on the calculation of electronic properties, and it is unknown how the presence of a long-lived metastable state influences the exact exchange-correlation potential as the particle number goes through $J_{max}$.  Although the accurate calculation of electron affinities has been explored with approximate DFT~\cite{LB10}, it has been unclear if one should use \emph{negative} electron affinities (as opposed to setting the affinity to zero) in the calculation of properties such as the chemical hardness which depend on derivatives of the energy with respect to particle number~\cite{CAPT11}.  If the behavior of the energy functional was known at fractional particle numbers above $J_{max}$, one could formulate answers to such questions.

In this work, we use complex ``densities'' calculated from complex-scaling theory~\cite{moiseyev1,reinhardt,simon} to study the derivative discontinuity in the Kohn-Sham exchange-correlation potential at the maximum number of bound electrons.  Since the complex energy functionals used give both the bound energies and metastable $(J_{max}+1)$ energies and lifetimes, the method presented here can be used to probe the behavior of the groud-state exchange-correlation potential around $J_{max}$.  

In previous work, we have demonstrated that the energy and lifetime of the lowest metastable state can be extracted from a complex density with a properly scaled energy functional~\cite{WW10, WM07}.  Also, a complex ``Kohn-Sham'' system can be constructed that facilitates self-consistent calculations on many-electron systems with the complex density as the primary variable~\cite{WW11}.  Zhang \emph{et al.} have developed a Levy-Nagy extension of this formalism to treat higher energy resonances (or excited states of the metastable system)~\cite{yu}, and in related work Maitra \emph{et al.} have considered autoionizing resonances within time-dependent DFT~\cite{KM09}.  Also, Ernzerhof and co-workers have developed an approach applicable to molecules connected to metallic leads where complex absorbing potentials are added within a complex-DFT framework \cite{E06, GEZ07}.  However, the complex potentials in the ``Density Functional Resonance Theory'' (DFRT) of Ref.~\cite{WW11} result from a variational calculation, and they are obtained self-consistently for the $N$-electron system treated as isolated, rather than added to the Hamiltonian from the start to model an open system.  

\section{Density Functional Resonance Theory and Complex Densities}
We use the complex analog of Kohn-Sham DFT, Density Functional Resonance Theory (DFRT), to treat metastable systems~\cite{WW11}.  The lowest-energy resonance energy and lifetime of a system is encoded into a corresponding complex resonance density~\cite{WW10} defined by,
\ben
n_\theta(\br)=\langle \Psi_\theta^L|\hat{n}(\br)|\Psi_\theta^R\rangle
\label{eq:den}
\een  
where $\hat{n}(\br)$ is the density operator, and $\langle\Psi_\theta^L|$ and $|\Psi_\theta^R\rangle$ are the left and right eigenvectors of the complex-scaled Hamiltonian, $\hat{H}_\theta$, corresponding to the Lowest-Energy Resonance, or LER (see Ref.~\cite{M11} for a review of complex-scaling theory). The angle $\theta$ is the complex-scaling angle in the transformation of the coordinates from $\vec{r}$ to $\vec{r} e^{i \theta}$.  We require that $n_\theta(\br)$ be normalized to the number of electrons, as real densities are:
\ben
\int d\br \ n_\theta(\br)=J
\label{eq:norm}
\een
The lifetime, $\cal{L}$, of the resonance is defined as $(2 \ \mbox{Im}(E_\theta))^{-1}$, where $E_\theta$ is a complex eigenvalue of $H_\theta$ corresponding to a pole in the scattering matrix.  The real part of $E_\theta$, $\cal{E}$, will be referred to as the ``resonance energy.''  

There is a one-to-one correspondance between complex-scaled external potentials and their corresponding LER complex densities~\cite{WM07}.  Therefore, the energy and lifetime of the LER can be extracted from $n_\theta$ with a properly scaled energy functional.  For $J$ electrons this functional is
\begin{eqnarray}
{\cal{E}}[n_\theta]-\frac{i}{2}{\cal{L}}^{-1}[n_\theta] &=& T_s^{\theta}[n_\theta]+\int d\br \  n_\theta(\br) v(\br e^{i\theta}) \nonumber \\ & & + E\H^{\theta}[n_\theta] + E\xc^{\theta}[n_\theta] 
\label{e:hypothesis}
\end{eqnarray}
in analogy to standard KS-DFT, with $T_s^\theta[n_\theta]=e^{-2i\theta} T_s[n_\theta]$  and $E\H^\theta[n_\theta]=e^{-i\theta}E\H[n_\theta]$, where $T_s[n_\theta]$ and $E\H[n_\theta]$ are the standard non-interacting kinetic energy and Hartree functionals evaluated at the complex densities.  Eq.~\ref{e:hypothesis} then defines $E\xc^\theta[n_\theta]$.  

The system of interacting electrons whose LER density is $n_\theta(\br)$ is mapped to one of $J$ particles 
moving independently in a complex ``Kohn-Sham'' potential $v_s^\theta(\br)$ defined such that its $J$ occupied complex orbitals 
$\{\phi_i^\theta(\br)\}$ yield the interacting LER-density via 
$n_\theta(\br)=\sum_{i=1}^J \langle\phi_i^{\theta,L}|\hat{n}(\br)|\phi_i^{\theta,R}\rangle$.  In Moiseyev's Hermitian representation \cite{M83}, the complex Kohn-Sham equations are:
\ben
\left(\begin{array}{cc}
\hat{h}_1 - \varepsilon_i & -\hat{h}_2 - 2\tau_i^{-1}\\
\hat{h}_2 + 2\tau_i^{-1} & \hat{h}_1 - \varepsilon_i\\
      \end{array}\right)
\left(\begin{array}{c}
{\mbox{Re}(\phi_i^\theta)}\\
{\mbox{Im}(\phi_i^\theta)}\\
      \end{array}\right)
= 0
~~,
\label{e:KS}
\een
where $\hat{h}_1 = -\half\cos(2\theta)\nabla^2 + \mbox{Re}(v_s^\theta(\br))$, and $\hat{h}_2=\half\sin(2\theta)\nabla^2+ \mbox{Im}(v_s^\theta(\br))$. The set of $\{\varepsilon_i\}$ and $\{\tau_i\}$ provide the orbital resonance energies and lifetimes of the
Kohn-Sham particles.  

The complex variational principle \cite{M11} along with the assumption that the orbitals used to construct the density can be expanded in an orthonormal basis leads to the Euler-Lagrange equation:
\ben
\frac{\delta E_\theta[n_\theta]}{\delta n_\theta}-\mu\int d\br n_\theta(\br) = 0~~.
\een 
Performing the variation in Eq.~\ref{e:hypothesis} and comparing with Eq.~\ref{e:KS} leads to an expression for the Kohn-Sham potential that is again analogous to that of standard KS-DFT:
\ben
v_s^\theta(\br)=v(\br e^{i\theta})+e^{-i\theta}v\H[n_\theta](\br)+v\xc^\theta[n_\theta](\br)~~,
\label{eqn:vs}
\een
where $v\xc^\theta[n_\theta](\br)=\delta E\xc^\theta[n_\theta]/\delta n_\theta(\br)|_{\rm LER}$.

\section{DFRT for Fractional Particle Numbers}
We start by considering the Levy-Lieb constained-search functional that has already been applied to both pure and ensemble states.
\begin{equation}
 F_{\sss LL}[n] = \underset{\hat{\Gamma} \rightarrow n}{\mbox{min}} \ \langle \hat{T} + \hat{V_{ee}} \rangle_{\Gamma}
\label{eq:levylieb}
\end{equation}
where $\hat{T}$ is the kinetic energy operator, $\hat{V_{ee}}$ is the electron-electron repulsion, and $\Gamma$ is a statistical mixture or ensemble of pure states, $\hat{\Gamma} = \sum_i |\Psi_i \rangle p_i \langle \Psi_i |$.  The sum of the probabilities $p_i$ is defined to be 1, and the expectation value of some observable is:
\begin{equation}
 \langle \hat{O} \rangle = \underset{i}{\sum} p_i \langle \Psi_i | \hat{O} | \Psi_i \rangle = \mbox{Tr} \left[ \hat{\Gamma} \hat{O} \right]
\end{equation}

For a fractional particle number $N$ between $J$ and $(J+1)$, the ensemble of a $J$-electron pure state and a $(J+1)$-electron pure state minimizes Eq.~\ref{eq:levylieb}.  This is true as long as the energy as a fraction of particle number obeys the ``concave-upward'' condition, or $E_J < (E_{J-1} + E_{J+1})/2$, for which no counter-example has been found in nature (but the condition has not been proven in general).  

In extending DFRT to fractional particle numbers, we define an ensemble of pure \emph{complex-scaled} states as:
\begin{equation}
 \hat{\Gamma}_\theta = \sum_i |\Psi_{\theta, i}^R \rangle p_i \langle \Psi_{\theta, i}^L |
\end{equation}
where $p_i$ has the same meaning and follows the same conditions as the usual, non-complex-scaled, case.  Then, the bi-expectation value is defined as
\begin{equation}
 O_\theta = \underset{i}{\sum} p_i \langle \Psi_{\theta, i}^L | \hat{O} | \Psi_{\theta, i}^R \rangle = \mbox{Tr} \left[ \hat{\Gamma}_\theta \hat{O} \right]
\end{equation}

We now make the assumption that the resonance of lowest energy is also the one with the longest lifetime. Like the convexity assumption above, this has never been proven, but it has been found empirically to be the typical case \cite{S73a,S73b}.
The constained-search functional
is then defined by:
\begin{equation}
 F_{\sss LL}^\theta[n_\theta] = \underset{\hat{\Gamma}_\theta \rightarrow n_\theta}{\mbox{min}}  \left( \begin{array}{c} \mbox{Re} \\ -2 \mbox{Im} \end{array} \right)  \mbox{Tr} \left[ \hat{\Gamma}_\theta (e^{-2 i \theta}\hat{T} + \hat{V}_{ee}^\theta) \right]
\label{eq:cslevylieb}
\end{equation}
where $\hat{V}_{ee}^\theta$ is the complex-scaled electron-electron interaction operator.  The ground-ensemble energy is:
\begin{equation}
 E^\theta[n_\theta] = \underset{n_\theta}{\mbox{min}} \left( \begin{array}{c} \mbox{Re} \\ -2 \mbox{Im} \end{array} \right) \left( F^\theta_{\sss LL}[n_\theta] +\int d\br \ n_\theta(\br) v(\br e^{i \theta}) \right)
\label{eq:engmin} 
\end{equation}  
If the specific statistical mixture is that of the $J$-electron state, $\Psi_{\theta,J}$, and the $(J+1)$-electron state, $\Psi_{\theta,J+1}$, the complex ground-ensemble energy for a fractional particle number $N$ ($J<N<J+1$) is,
\begin{equation}
 E^\theta(N) = [1-(N-J)] E_J^\theta + (N-J) E_{J+1}^\theta
\end{equation}
where $E_J^\theta$ is the bound ground state (or if no bound state exists, lowest metastable state) complex energy of the $J$-electron system, and likewise $E_{J+1}^\theta$ is the bound ground state (or if no bound state exists, lowest or second-lowest metastable) complex energy of the $J+1$ electron system.  Again, the lifetime, $\cal{L}$, of the state is given by $(2 \ \mbox{Im}(E_\theta))^{-1}$ (this will be zero if the state is bound), and the real part of $E_\theta$, $\cal{E}$, is the ground-state energy if the system is bound or the ``resonance energy'' if the system is unbound.  The complex ground-ensemble density is,
\begin{equation}
 n_\theta^N (\textbf{r}) = \left[ 1-(N-J) \right] n_\theta^J (\textbf{r}) + (N-J)n_\theta^{J+1} (\textbf{r}) 
 \label{eq:denens}
\end{equation}
One might ask whether the ensemble of the $J$ and $(J+1)$ electron states minimizes Eq.~\ref{eq:engmin}, or whether we have a ``concave-upward'' condition given that we now are dealing with complex energies.  The real part of the energy (position of the resonance or bound state) follows the concave-upward condition already discussed.  However, assuming that the LER has the longest lifetime, the imaginary part of the energy as a function of particle number would be concave-down (at least up to $J_{max}+1$).  Yet, the constrained search in Eq.~\ref{eq:engmin} is defined as giving the \emph{minimum} width, $\Gamma$, or the \emph{maximum} lifetime, and the width does follow a ``concave-upward'' condition.

This extension of DFRT allows us to study the derivative discontinuity at integer values below $J_{max}$, because DFRT can be applied to bound states, but it also allows us to investigate the discontinuity (if there is one) at $J_{max}$.

\section{Significance of the Complex Kohn-Sham Orbital Energies}
One related topic that needs to be discussed before studying the discontinuity at $J_{max}$ is the physical meaning of the Kohn-Sham orbital energies.  One can do this by looking at the assymptotic behavior of densities.  If we consider the tails of a real, bound ground-state density, we know that they decay exponentially like $e^{-2 \sqrt{2 I}r}$, where $I$ is the ionization potential~\cite{AB85, KD80}.  The highest occupied Kohn-Sham orbital energy is then given by the negative of $I$. However, in DFRT we must take careful note of how a resonance wavefunction, $\phi$, decays.   As $r \rightarrow \infty$,
\begin{equation}
 \phi_{\alpha,n} (x) \longrightarrow C e^{i k_{\alpha,j} r}
\end{equation}
where,
\begin{equation}
 k_{\alpha,j} = \sqrt{2 (E_\alpha - E_j^{th})}
\label{eq:k}
\end{equation}
And the complex-scaled density would decay as:
\begin{equation}
 n_{\alpha,j}^\theta (x) \longrightarrow C^2 e^{i 2 k_{\alpha,j} r e^{i \theta}}
\label{eq:den2}
\end{equation}
In these expressions the index $\alpha$ defines the eigenvalue of the complex-rotated Hamiltonian, $j$ labels the decay channel, and $C$ is a constant~\cite{GGM10}.  $E_j^{th}$ is the threshold energy of the decay channel, and $E_\alpha$ is the complex eigenvalue of the resonance.  The wavefunction decay is governed by the energy difference $(E_\alpha - E_j^{th})$.  

For simplicity of discussion, consider the LER of a $(J+1)$-electron unbound system that decays to a bound $J$-electron system (this is the single dominant channel of decay).  $E_j^{th}$ would equal the energy of the $J$-electron state ($E_{J}$).  Therefore the tail of the complex density of the resonance behaves like $\mbox{exp} \left( i 2 \sqrt{2 (E_{(J+1)} - E_{J})} r \right)$, where $E_{(J+1)}={\cal{E}}_{(J+1)} - \frac{\Gamma}{2}i$ is the complex LER resonance eigenvalue (${\cal{E}}_{(J+1)}$-position or energy, $\Gamma$-width or inverse lifetime).  The energy difference in the exponential tail of the complex density can then be related to the NEA of the $J$-electron system, $A=(E_{J}-{\cal{E}}_{(J+1)})$:
\begin{equation}
 (E_\alpha - E_j^{th})=({\cal{E}}_{(J+1)} - \frac{\Gamma}{2}i - E_{J}) = (-A - \frac{\Gamma}{2}i)~~,
\label{eq:aff}
\end{equation}

For a $(J+1)$-electron unbound system with multiple decay channels (or a non-sequential decay process) it is useful to look at the time-dependence of the states (partial widths and branching ratios) to gain understanding of the decay process~\cite{GGM10}.  The relation to a NEA is not immediately clear.  

The complex density of the LER in DFRT, constructed from the KS orbitals, behaves exactly as in Eq.~\ref{eq:den2}, with the wavenumber given in terms of \emph{Kohn-Sham} quantities: $k_{\alpha,j} = \sqrt{2 (\epsilon_{\sss H}^\theta - \epsilon_j^{th})}$.  $\epsilon_j^{th}$ is the KS ``threshold energy,'' and $\epsilon_{\sss H}^\theta$ is the complex HOMO resonance energy.  For the system with one decay channel discussed above, comparing with Eq.~\ref{eq:k},
\begin{equation}
 (\epsilon_{\sss H}^\theta - \epsilon^{th}) = (-A - \frac{\Gamma}{2}i)~~,
\label{eq:hmt}
\end{equation}
This provides some physical interpretation for $\epsilon_{\sss H}^\theta$, but we need a clear definition of $\epsilon_j^{th}$.  In analogy to ground-state DFT, one can write the LER's complex energy in terms of a sum of KS complex orbital energies~\cite{WW11}:
\begin{eqnarray}
{\cal{E}}_{(J+1)} [n_\theta] - \frac{\Gamma [n_\theta]}{2}i &=& \epsilon_{\sss H}^\theta + \sum_{i=1}^{J} \epsilon_{i}^\theta +E\Hxc^\theta[n_\theta] \nonumber \\
 & & -\int d\br v\Hxc^\theta (\br) n_\theta(\br)
\end{eqnarray}
Where the HOMO energy has been pulled out from the eigenvalue sum.  We can then write the following expression for $\epsilon_{\sss H}^\theta$,
\begin{equation}
 \epsilon_{\sss H}^\theta  =  {\cal{E}}_{(J+1)} [n_\theta] - (\Gamma [n_\theta]/2)i - \xi [n_\theta] 
\label{eq:homo}
\end{equation}
where,
\begin{equation}
 \xi [n_\theta]  = \sum_{i=1}^{J} \epsilon_{i}^\theta +E\Hxc^\theta[n_\theta] -\int d\br v\Hxc^\theta (\br) n_\theta(\br)
\label{eq:xi}
\end{equation}
Comparing Eq.~\ref{eq:homo} with Eq.~\ref{eq:hmt},
\begin{eqnarray}
 \epsilon^{th} &=& {\cal{E}}_{(J+1)} [n_\theta] - (\Gamma [n_\theta]/2)i - \xi [n_\theta] + (A + \frac{\Gamma}{2}i) \nonumber \\
  &=& E_{J} - \xi [n_\theta] 
 \label{eq:the}
\end{eqnarray}
Therefore, $\epsilon^{th}$ is just the threshold energy in the real interacting system, $E_{J}$, shifted by $\xi$. Now with this definition of $\epsilon^{th}$ we can relate the HOMO energy of DFRT to physical quantities via Eq.~\ref{eq:hmt} (for a system with a single dominant decay channel).  Also, note that the standard Koopmans' theorem for DFT is recovered when the system of interest is bound.

Table~\ref{tab:orbsumm} summarizes this analysis.

\begin{table*}
\begin{center}
\begin{tabular}{lll}
System: & Decay Channel: & $\epsilon_{{\sss H}}^\theta$   \\
\hline
Bound $J$-electron & No decay & $-I$ \\
Metastable $(J+1)$-electron \quad & $J$ electron bound state \quad & $(-A - \frac{\Gamma}{2}i) + \epsilon^{th}$ \\
Metastable $(J+1)$-electron \quad & Multiple & Dependent on partial widths / branching ratios \\
\hline
\multicolumn{3}{l}{} \\
\multicolumn{3}{l}{$\epsilon_{{\sss H}}^\theta$, KS-DFRT HOMO energy} \\
\multicolumn{3}{l}{$I$, Positive ionization potential of a $J$-electron system} \\
\multicolumn{3}{l}{$\epsilon^{th}$, KS ``threshold energy'' defined in Eq.~\ref{eq:the}} \\
\multicolumn{3}{l}{$A$, Negative electron affinity of a $J$-electron system} \\
\multicolumn{3}{l}{$\Gamma$, $(J+1)$-electron resonance width (inverse lifetime)} \\
\end{tabular}
\end{center}
\caption{Summary of the significance of Kohn-Sham DFRT orbital energies with definitions of the relevant quantities.}
\label{tab:orbsumm}
\end{table*}

\section{Model Problem}
To study the behavior of the exchange-correlation potential at $J_{max}$ we consider a system of two interacting electrons moving in a one-dimensional potential which can be solved exactly using finite difference methods.  We study a Hamiltonian where the electrons interact via a soft-Coulomb potential of strength $\lambda$:
\ben
\hat{H}=\sum_{i=1}^2\left[-\half\frac{d^2}{dx_i^2}+v(x_i)\right]+\frac{\lambda}{\sqrt{1+(x_1-x_2)^2}}~~,
\label{e:Hamiltonian}
\een
with,
$v(x) = a\left[\sum\limits^{2}_{j=1} \left(1+e^{-2 c (x + (-1)^jd)}\right)^{-1} \right]-\alpha e^{-\frac{x^2}{b}}$. 
where $a$, $\alpha$, $b$, $c$, and $d$ are constants.  
Note that the exact complex energies are theoretically independent of the scaling parameter $\theta$, but numerical approximation introduces some $\theta$-dependence.  This dependence dissapears in the infinite basis-set limit, but it can also be addressed with a finite basis set or finite grid by calculating ``$\theta$-trajectories''~\cite{M11} for an optimum $\theta$.  This procedure was used, but for the above model problem we were able to use enough grid points to extinguish most of the $\theta$-dependence of the energies.  The ensemble density, Eq.~\ref{eq:denens}, for 1- and 2-electron systems can be inverted to construct the exact complex KS potential:
\begin{equation}
 v_{s}^\theta(x) = e^{-2 i \theta} \frac{\nabla^2 \sqrt{n_\theta^N(x)}}{2 \sqrt{n_\theta^N(x)}} + \epsilon_{\sss H}^\theta~~,
\label{eq:invert}
\end{equation}
where $\epsilon_{\sss H}$ is the highest occupied Kohn-Sham orbital.  The Hartree contribution to the KS potential can be found directly from the density,
\begin{equation}
 v\H^\theta [n_\theta^N] (x) = \int dx \ v_{ee}(x'e^{i \theta},x e^{i \theta}) n_\theta^N (x')
\end{equation}
The exact exchange-correlation potential can then be found by the difference,
\begin{eqnarray}
 v\xc^\theta [n_\theta^N] (x) &=&  e^{-2 i \theta} \frac{\nabla^2 \sqrt{n_\theta^N(x)}}{2 \sqrt{n_\theta^N(x)}} - v\H^\theta [n_\theta^N] (x)  \nonumber \\
 & & - v(x e^{i \theta}) + \epsilon_{\sss H}^\theta
\end{eqnarray}
For the bound case and for particle numbers between 0 and 2, Sagvolden and Perdew proved that the kinetic and Hartree parts of this potential change continuously as the particle number crosses an integer~\cite{SP08}.  Note that these potentials depend on $\theta$, so we only compare potentials calculated with the same scaling parameter~\cite{WW11}.  For certain values of the parameters this potential can be made to support 2 bound states.  We show that in this case $v_{\xc}^\theta$ exibits discontinuous jumps for integers $J < J_{max}$.  For a different set of parameters, $v(x)$ can be made to have a very weakly bound 1-electron state and a metastable but long-lived 2-electron state.  For such a system, the maximum number of bound electrons, $J_{max}$, is one, and the system has a negative electron affinity, $A= {\cal{E}}_{J_{max}} - {\cal{E}}_{J_{max}+1}$.  Calculations on the bound 1-electron and unbound 2-electron states then allows $v_{\xc}^\theta$ to be examined for the integer $J=J_{max}$.

\section{Results for $J < J_{max}$}
\begin{figure}
  \scalebox{0.26}{\includegraphics{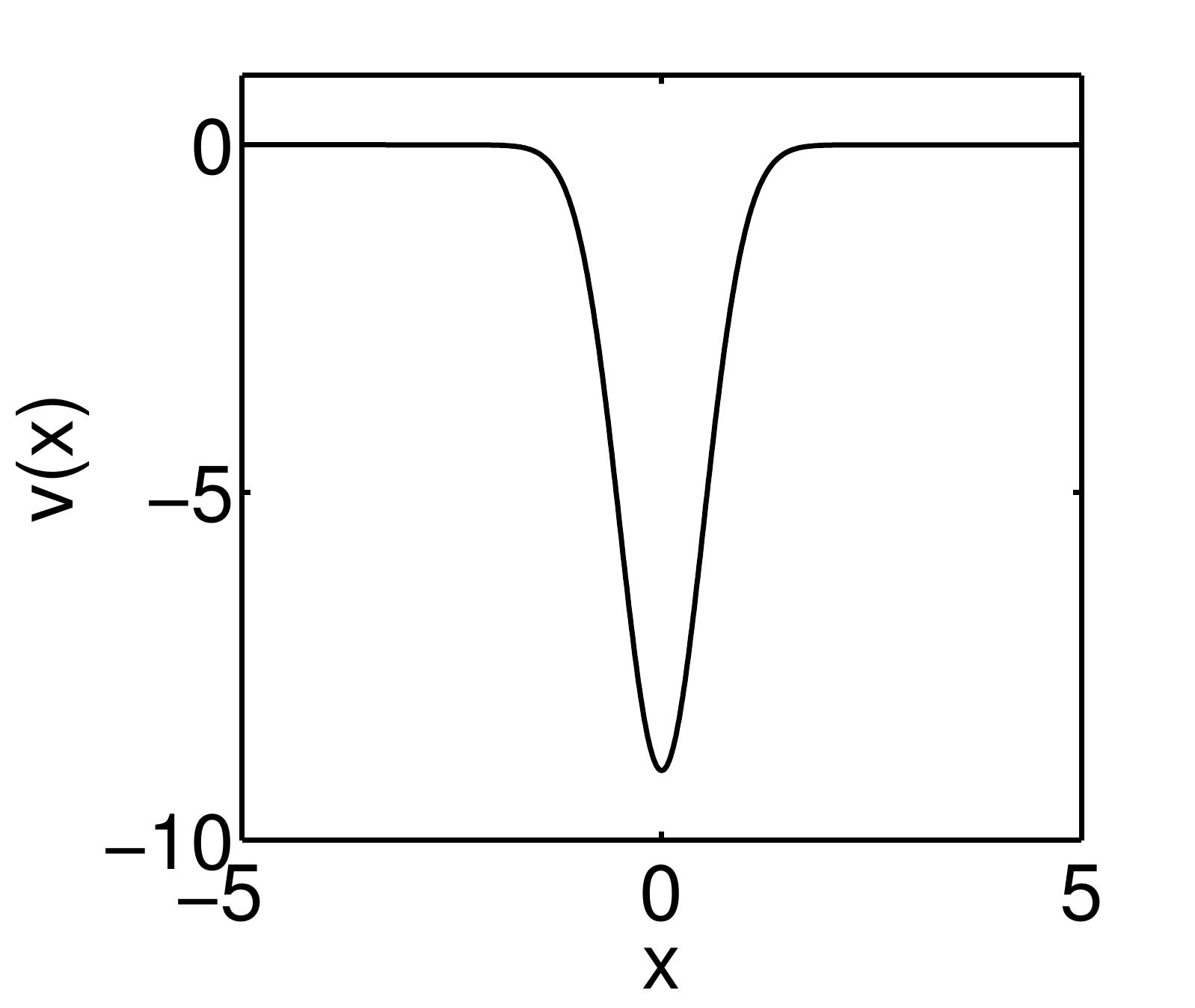}} 
  \scalebox{0.25}{\includegraphics{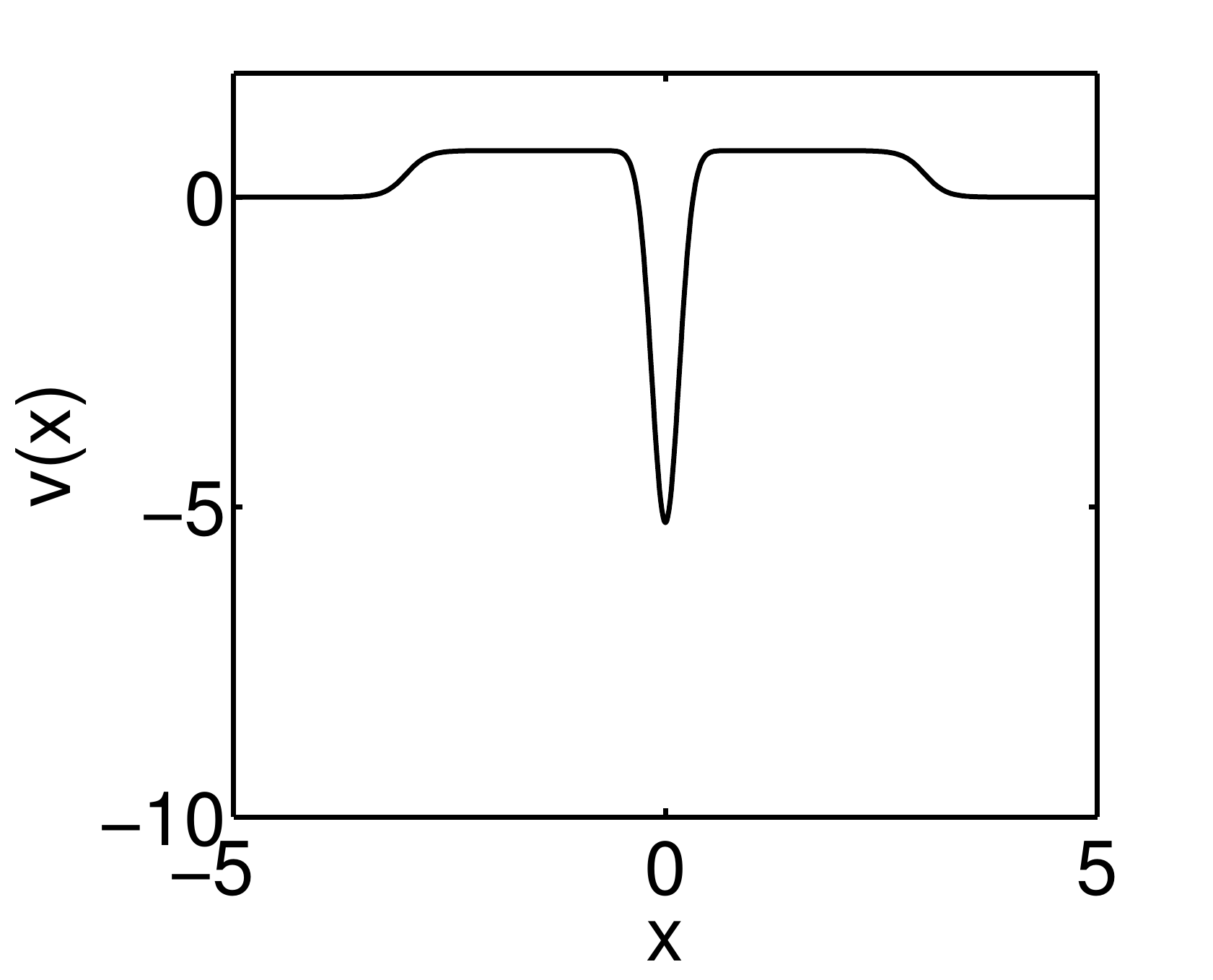}}
  \caption{The two model potentials used in this study.  On the left, $v(x)$ with $a=0$, $\alpha=9$, $b=0.5$, $c=0$, and $d=0$. This potential has both a 1 electron and 2 electron bound state (when $\lambda=1$ in $v_{ee}$). On the right, $v(x)$ with $a=0.75$, $\alpha=6$, $b=0.05$, $c=4$, and $d=3$. This potential has a very weakly bound 1 electron state and a metastable but long-lived 2 electron state (when $\lambda=1$ in $v_{ee}$).}
  \label{fig:pots}
\end{figure}

\begin{figure}
  \scalebox{0.25}{\includegraphics{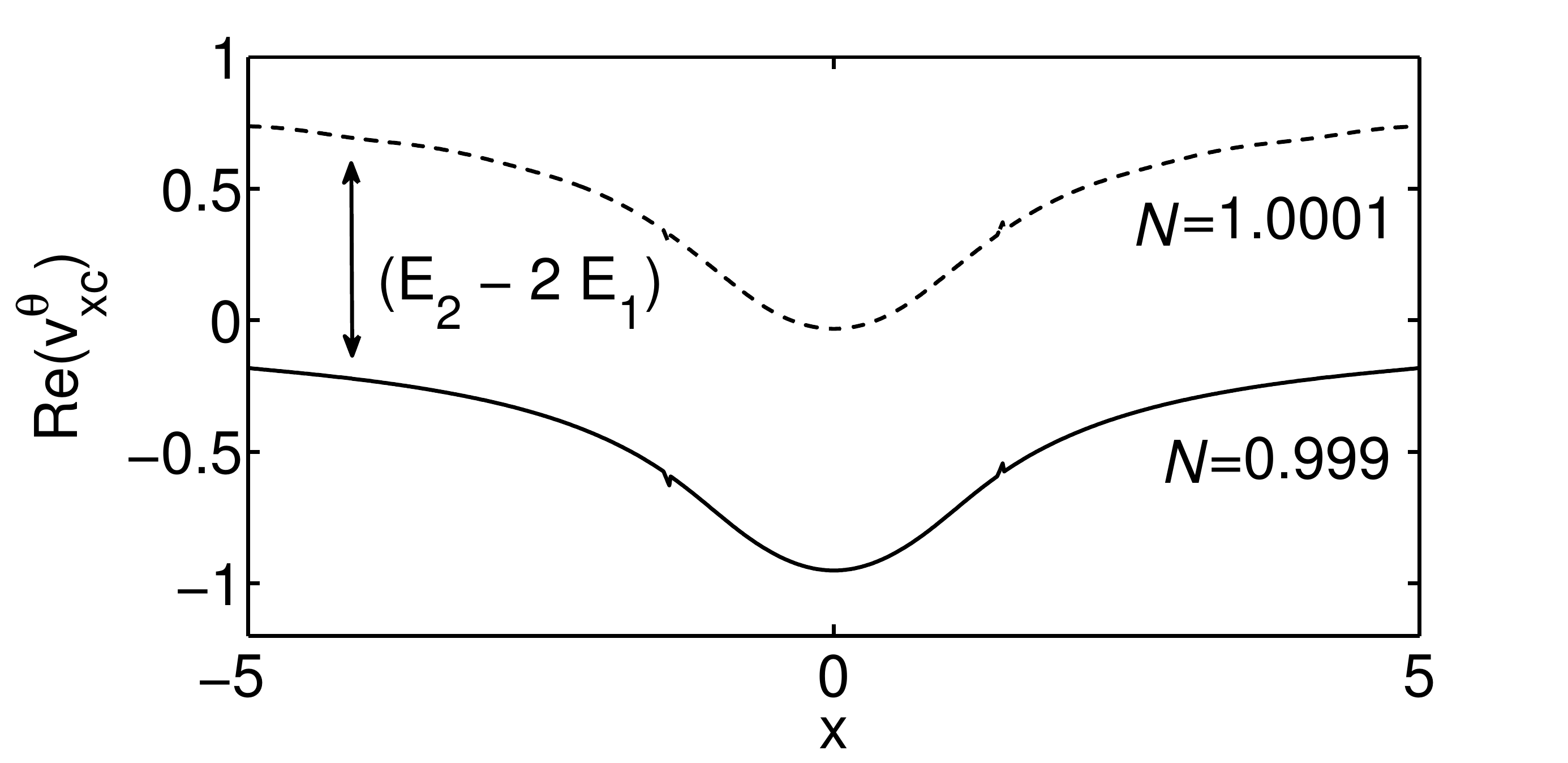}} \\
  \scalebox{0.26}{\includegraphics{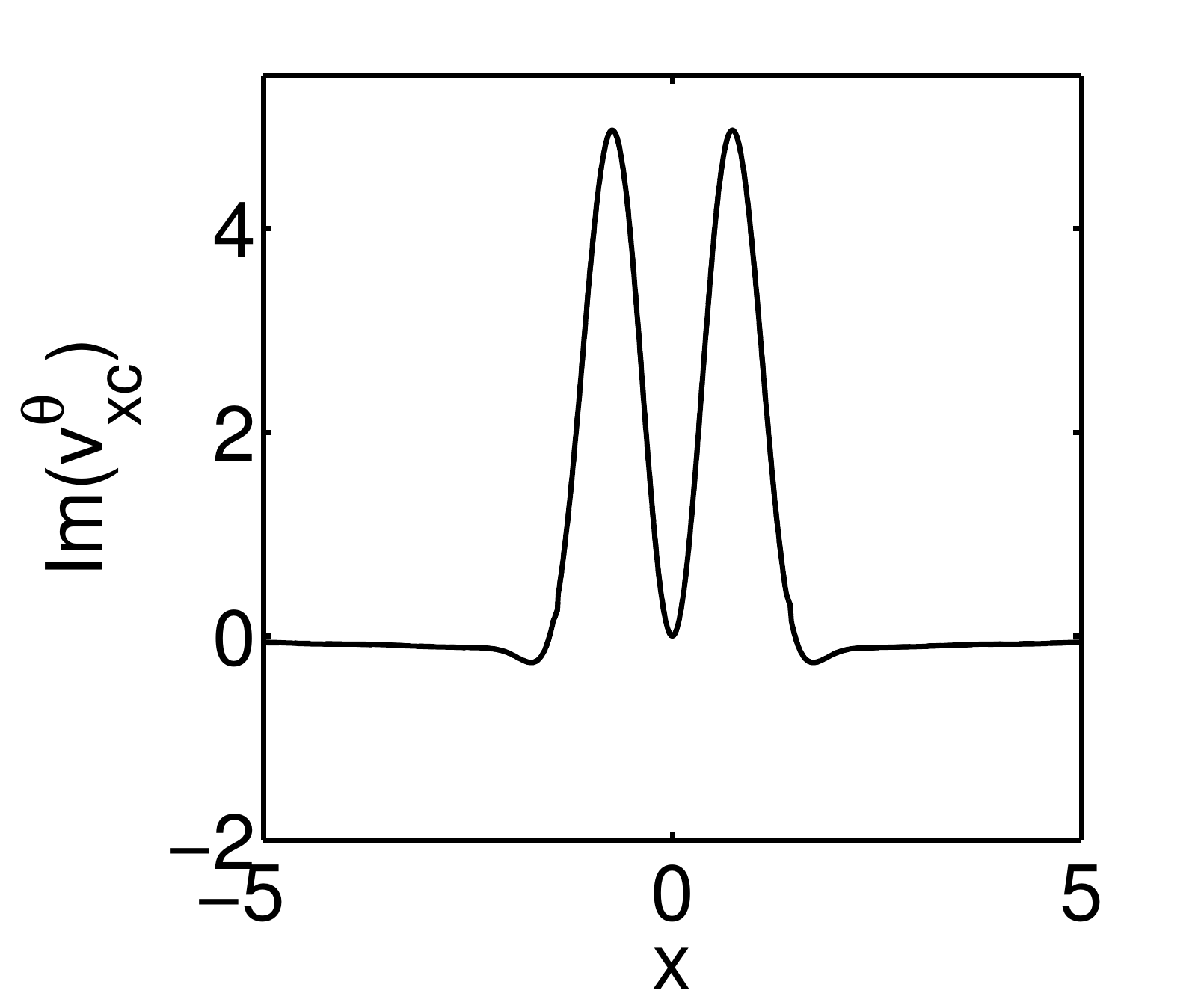}} 
  \scalebox{0.26}{\includegraphics{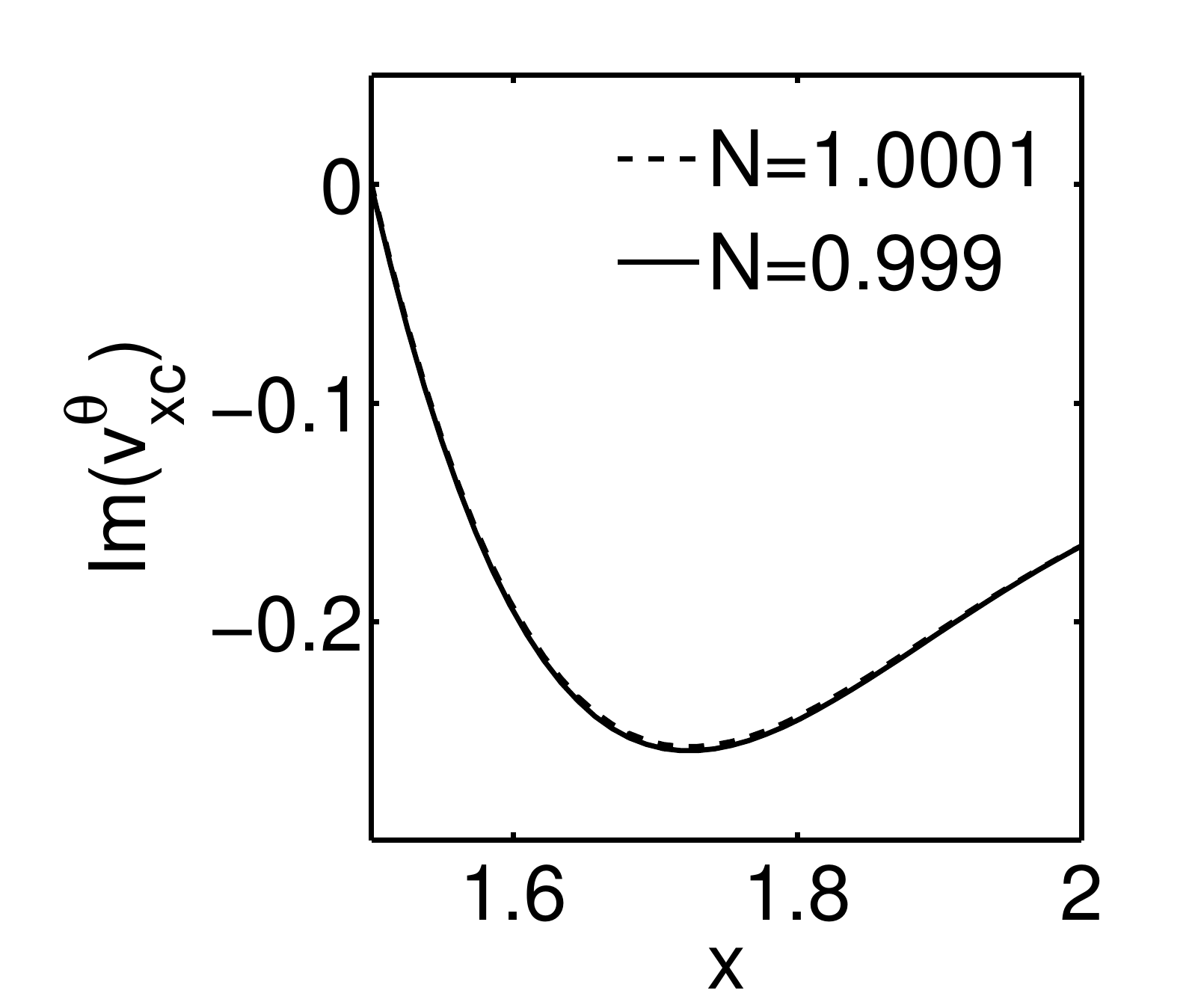}}
  \caption{The real and imaginary parts of $v_{\xc}^\theta$ above and below the integer $J=1$ when $J_{max}=2$ ($\theta=0.35$). The real part of the potential jumps by $E_2 - 2E_1$, as in ensemble DFT. The imaginary part does not experience any discontinuity (solid and dashed lines are on top of each other in the bottom panels).}
  \label{fig:jdis}
\end{figure}

The model potential $v(x)$ with $a=0$, $\alpha=9$, $b=0.5$, $c=0$, and $d=0$ and $v_{ee}$ with $\lambda=1$ supports both a 1-electron bound state and a 2-electron bound state.  The external potential is shown in Fig.~\ref{fig:pots} (left panel).  The 1-electron bound energy is $E_1 = -6.38$, and the 2-electron bound energy is $E_2 = -11.84$.
Both the 1- and 2-electron complex densities are found exactly with finite differences.  The ensemble density is formed according to Eq.~\ref{eq:denens} and the exact $v_{\xc}^\theta$ is calculated for different values of $N$.  This model system is similar to the one studied in Ref.~\cite{SP08}.  In that work, the purely real XC potential was shown to experience a jump at the integer $J=1$.  We find a similar result when using our complex-density analysis.  The change in the complex density for a value of $N$ infinitesimally less than or greater than $J=1$ is negligible, yet the chemical potential, or Kohn-Sham HOMO energy (see Table~\ref{tab:orbsumm}), experiences a purely real jump of $\Delta \mu = E_2 - 2 E_1$ on either side of $J=1$.  Also, since there is a smooth change in the complex density, the change in the Hartree potential is negligible on either side of $J=1$.  Therefore, the complex XC potential must compensate for this shift in the chemical potential by a positive jump (see Figure~\ref{fig:jdis}).  
There is no jump in the imaginary part of the XC potential because there is no jump in the imaginary part of $\mu$ or the HOMO energy.  In other words, both the $J=1$ and $J=2$ states are bound and their energies have an imaginary part of exactly zero.

\section{Results at $J = J_{max}$}
\begin{figure}
  \scalebox{0.25}{\includegraphics{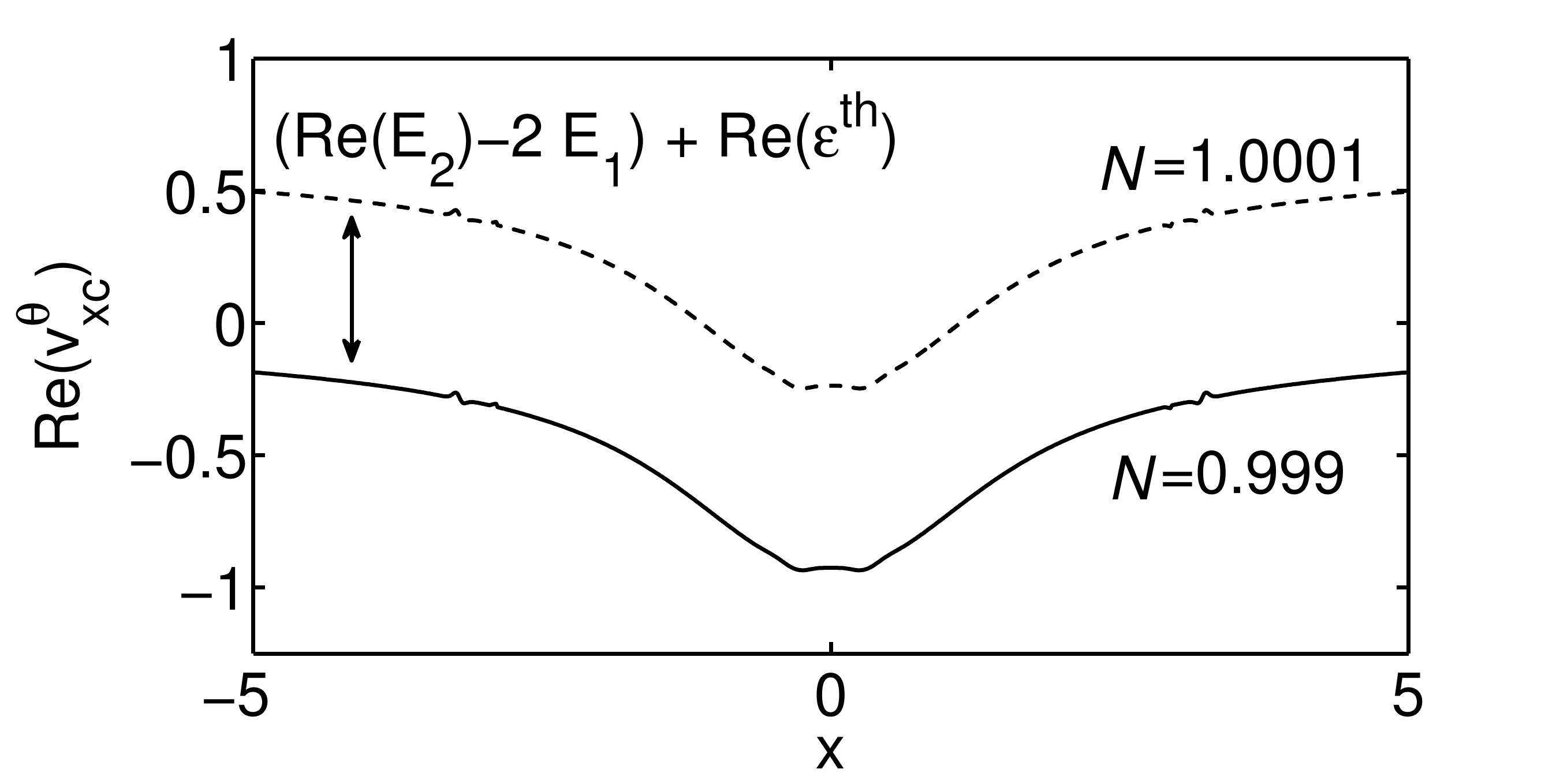}} \\
  \scalebox{0.25}{\includegraphics{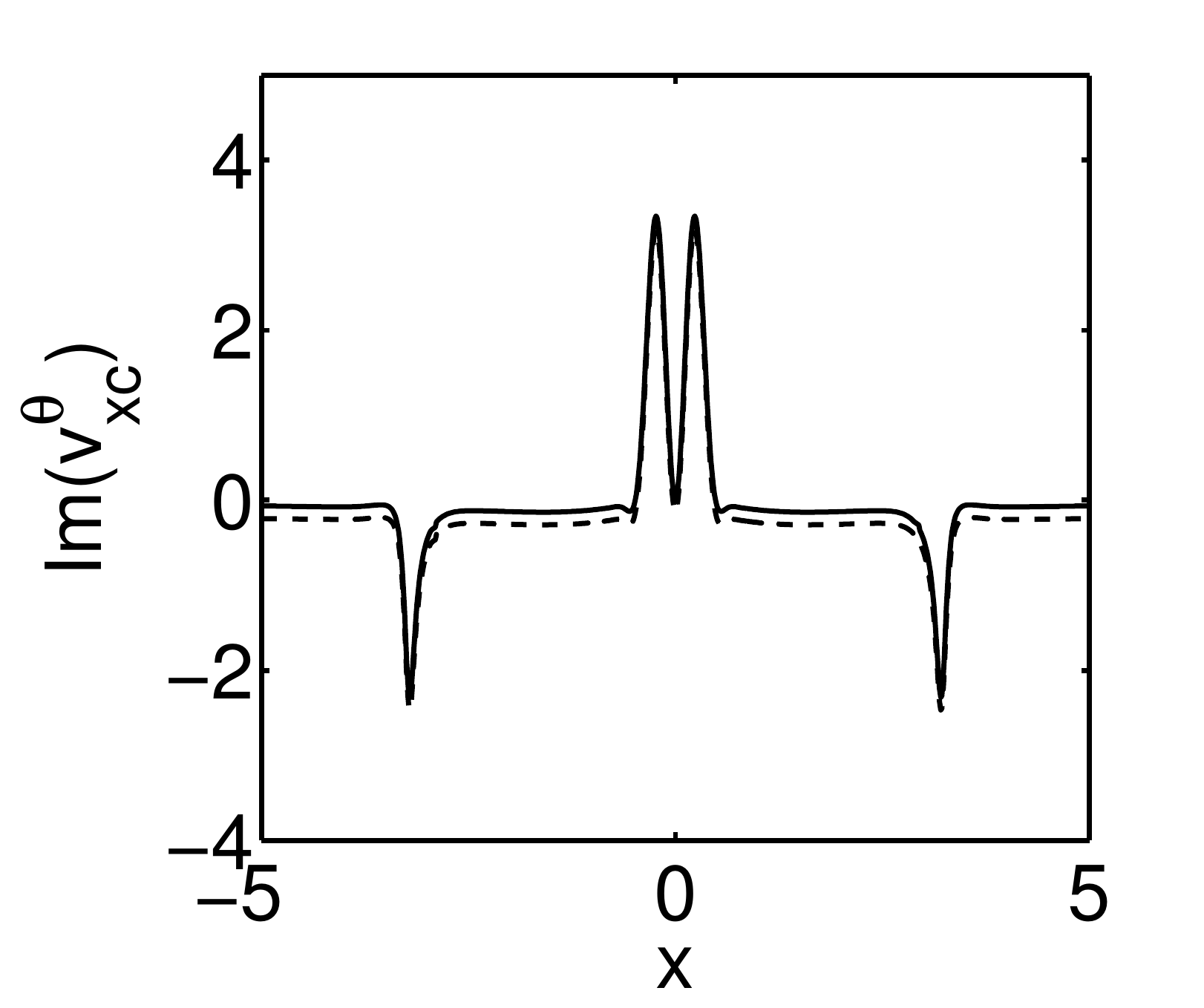}} 
  \scalebox{0.25}{\includegraphics{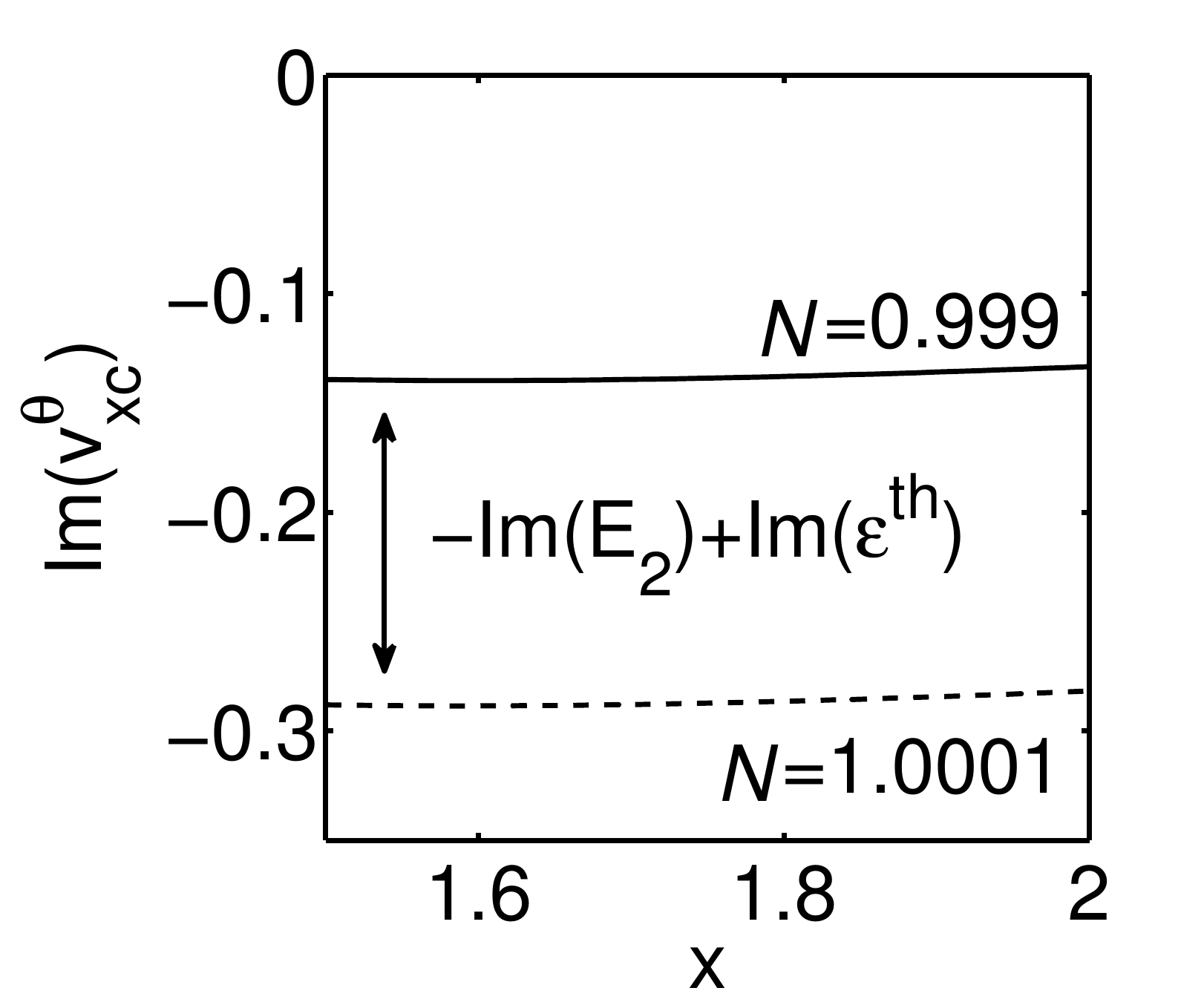}}
  \caption{The real and imaginary parts of $v_{\xc}^\theta$ above and below the integer $J=1$ when $J_{max}=1$ ($\theta=0.35$). It is clear hear that the real part jumps by $(\mbox{Re}(E_2) - E_1)$, but we must take a closer look at the imaginary part to see the discontinuous jump of $-\Gamma/2 = \mbox{Im}(E_2)$.}
  \label{fig:jmaxdis}
\end{figure}

The model potential $v(x)$ with $a=0.75$, $\alpha=6$, $b=0.05$, $c=4$, and $d=3$ and $v_{ee}$ with $\lambda=1$ supports only one bound state with 1-electron and one metastable state with 2 electrons.  This potential is shown in Fig.~\ref{fig:pots} (right panel).  Its steps mimic the centrifugal barriers present in 3D potentials which give rise to shape resonances.  The 1-electron bound energy is $E_1 = -0.86$, and the 2-electron complex energy of the metastable state is $E_2 = -0.63 - 0.066 i$.  Both the bound 1-electron and metastable 2-electron complex densities are found exactly with finite differences.  The ensemble density is formed according to Eq.~\ref{eq:denens} and $v_{\xc}^\theta$ is calculated for different values of $N$.  Fig.~\ref{fig:jmaxdis} shows both the real and imaginary parts of $v_{\xc}^\theta$ above and below $J_{max}=1$.  The structure of these potentials is similar to the bound potentials in Fig.~\ref{fig:jdis} except for the dramatic scars in the imaginary part from the potential steps which influence the lifetime of the resonance.  
From Fig.~\ref{fig:jmaxdis} one can see that $v_{\xc}^\theta$ experiences a discontinuous jump in its real part, and if we zoom in on $v_{\xc}^\theta$ a jump in its imaginary part also becomes visible (see Fig.~\ref{fig:jmaxdis}).  These jumps are to compensate for the complex shift in the chemical potential.  The HOMO energy just below $J_{max}=1$ is $\epsilon_{{\sss H},1}=E_1$, but the HOMO energy just above $J_{max}=1$ is $\epsilon_{{\sss H},2}= (-(E_1-{\cal{E}}_2) - \frac{\Gamma}{2}i) + \epsilon^{th}$, where $\Gamma$ is the width of the 2-electron resonance (see Table~\ref{tab:orbsumm}). For this system, we can calculate the exact $\epsilon^{th}$ according to Eq.~\ref{eq:the}.  Then $\epsilon_{{\sss H},2}=-0.17-0.15 i$.  Therefore, $\Delta \mu = (-(E_1-{\cal{E}}_2) - \frac{\Gamma}{2}i) + \epsilon^{th} - E_1 = 0.69-0.15 i$.  Note that here the constant jump in the real part of the XC potential is positive, as in previous studies, but the jump in the imaginary part is negative because of the negative shift in the chemical potential from the existence of a long-lived metastable state.  

\begin{figure}
  \scalebox{0.32}{\includegraphics{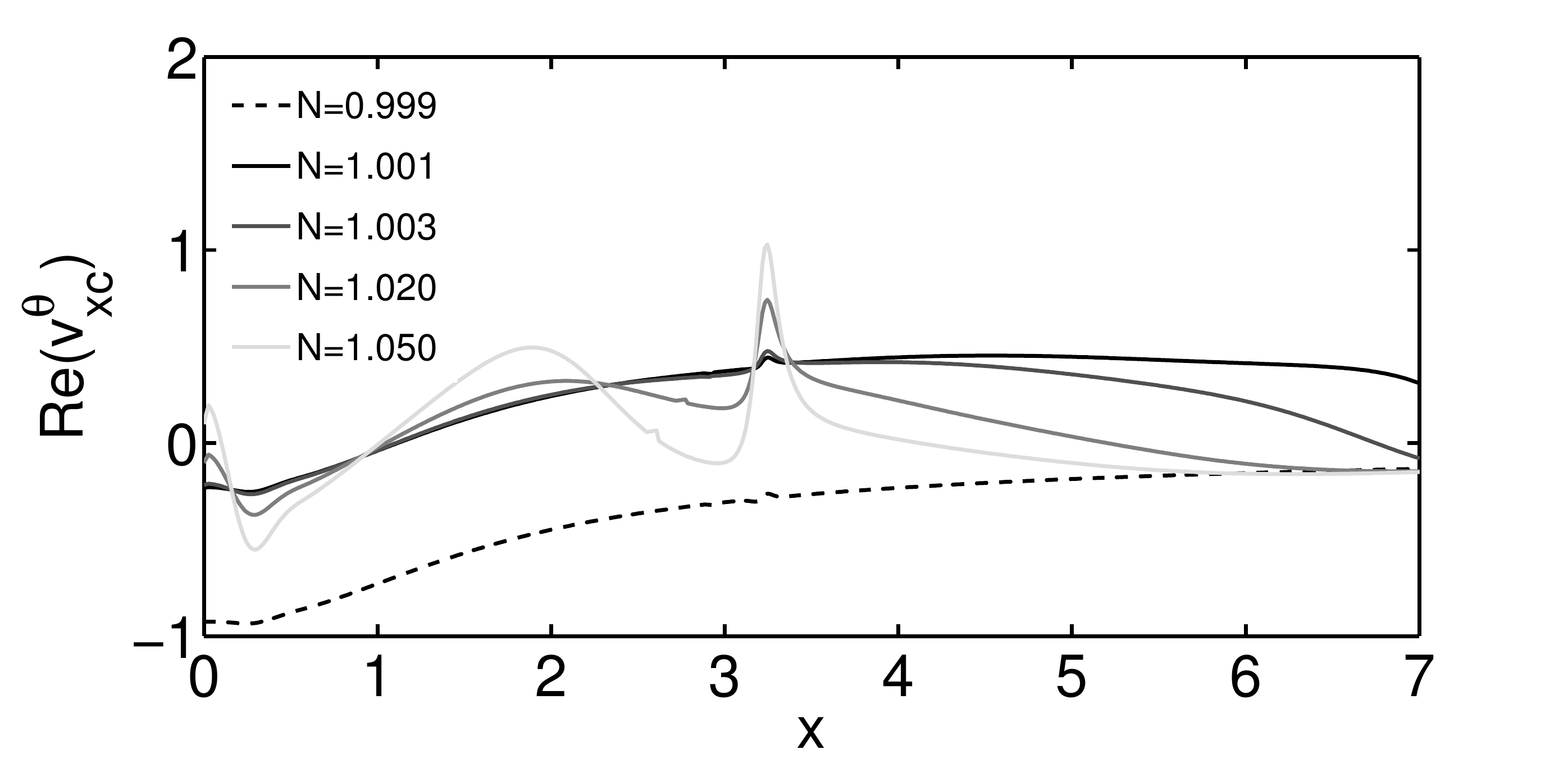}} 
  \caption{Asymptotic behavior of $v_{\xc}^\theta$ as particle number increases ($\theta=0.35$).}
  \label{fig:decay}
\end{figure}

In the previous work by Sagvolden and Perdew~\cite{SP08}, it was observed that the exact $v\xc$ of ground-state DFT slightly above an integer was shifted upward from the $v\xc$ slightly below an integer.  This constant upward shift extended out to a radius that depended on $N$, the fractional number of particles, and this radius extends further out as $N$ approaches the integer from above.  We can see the same type of asymptotic behavior in the real part of $v_{\xc}^\theta$ as a function of the fractional particle number, $N$ (see Fig.~\ref{fig:decay}).  As $N \rightarrow J_{max}^+$, the radius out to which $\mbox{Re}(v_{\xc}^\theta)$ is shifted upward by a constant goes to infinity.  Although this radius goes to infinity as $N$ gets infinitesimally close to $J_{max}$ from above, there is no contradiction with the exact condition stating that the Kohn-Sham potential should go to zero at infinity.  As stated in Ref.~\cite{SP08,PL83}, if $(N-J)$ is finite there will always be a radius at which the constant shift falls off.  For larger values of $N$ the shift is clearly not a constant, instead $v_{\xc}^\theta$ shows a rich structure resulting from the transient nature of the $(J_{max}+1)$-electron state.  

\begin{figure}
  \scalebox{0.29}{\includegraphics{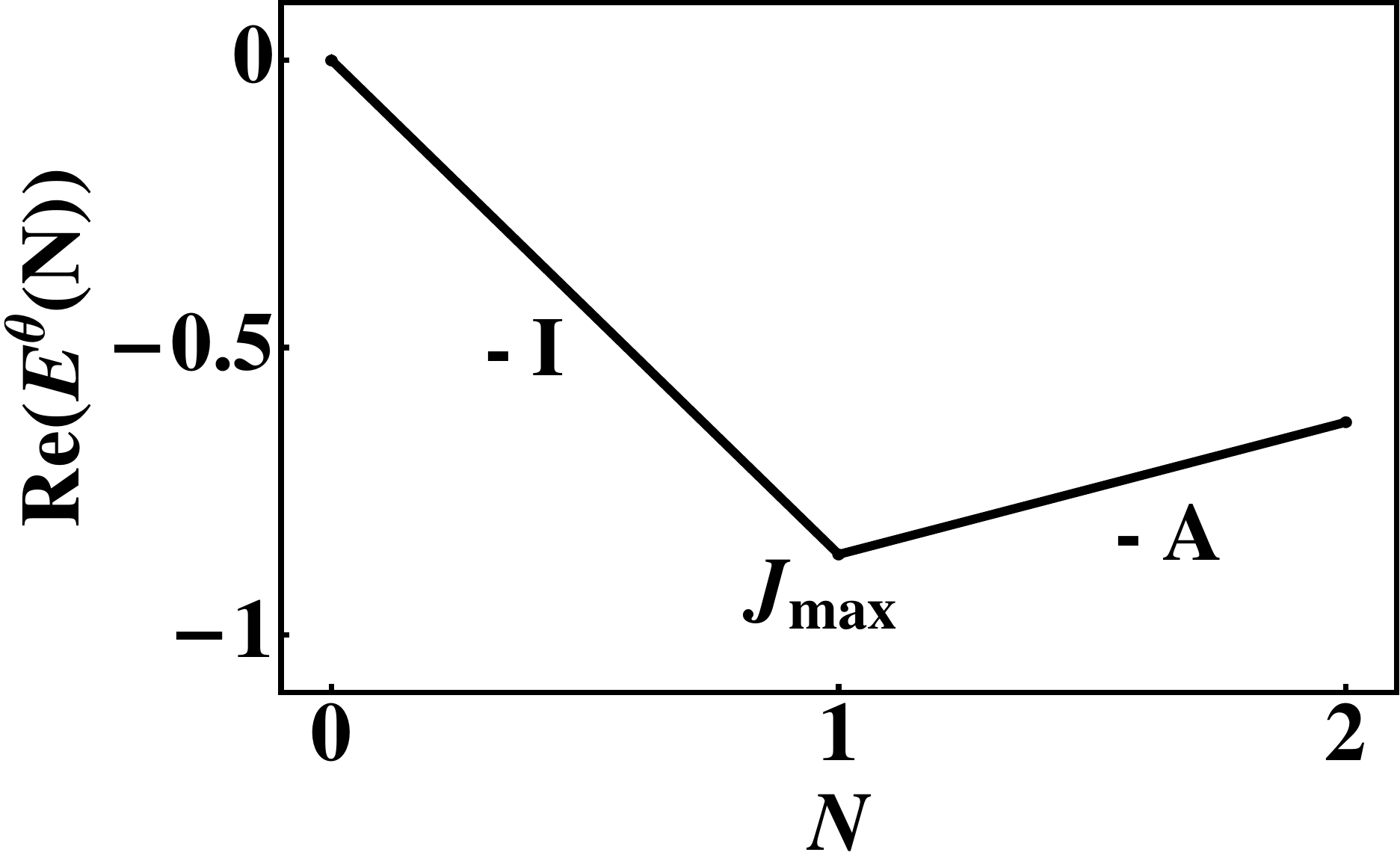}} \\
\vspace{15pt}
  \scalebox{0.272}{\includegraphics{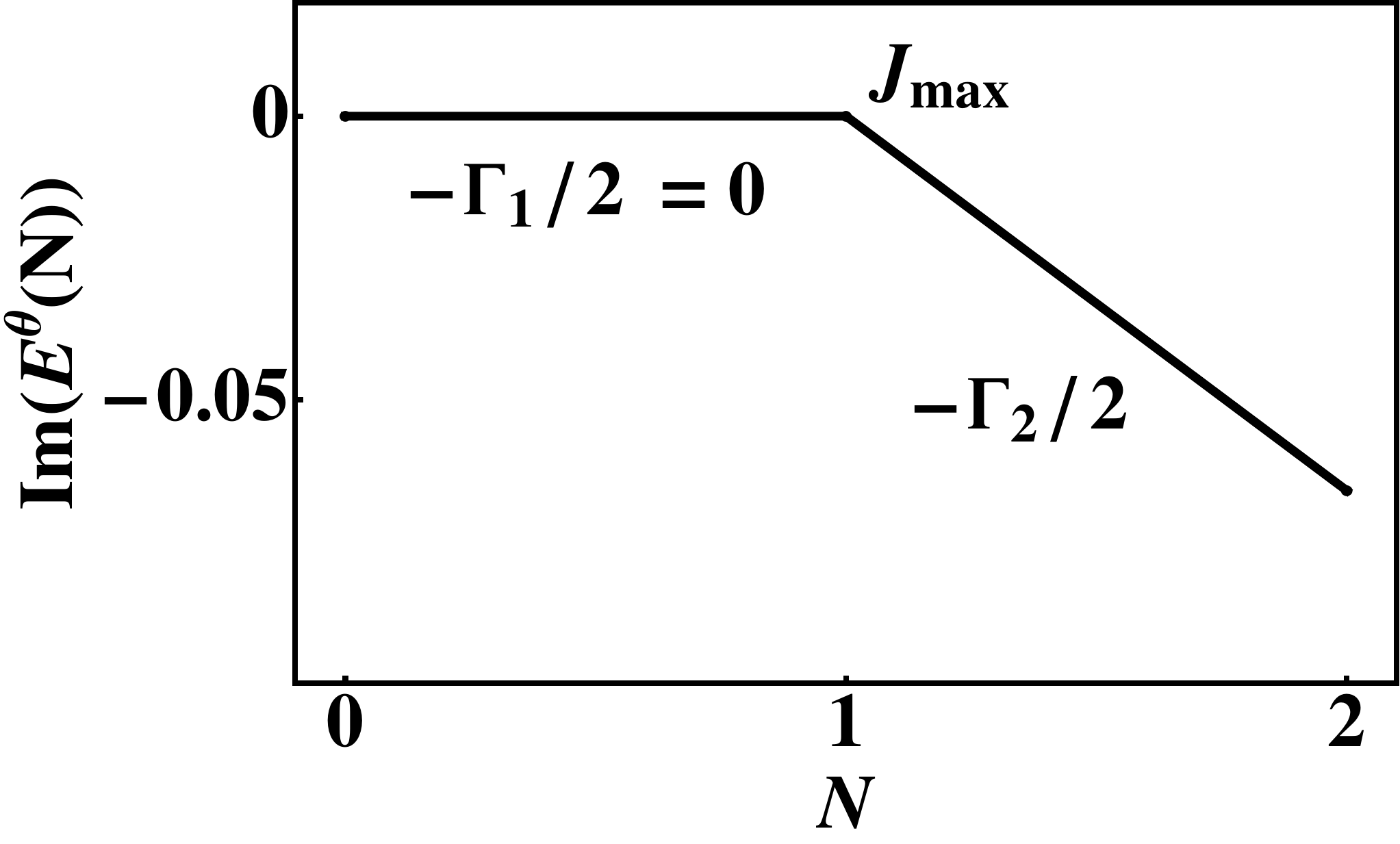}} 
  \caption{The real part of an open system's energy as a function of particle number experiences discontinuities at integer values of particle number including $J_{max}$.  The imaginary part experiences discontinuities at integer values of particle number greater than of equal to $J_{max}$.  This example is for the model system with $J_{max}=1$. ($I$ is the ionization potential of the 1 electron system, $A$ is the affinity of the 1 electron system, $\Gamma_1$ is the resonance width of the 1 electron system, and $\Gamma_2$ is the width of the 2 electron system)}
  \label{fig:dis}
\end{figure}

When one electron is bound and two electrons are unbound the energy as a function of particle number can be represented by two piecewise linear graphs.  First, a graph that is similar to the cartoon (Fig.~\ref{fig:cartoon}) in the introduction (see Fig~\ref{fig:dis}) shows that the real part of the energy of an open system experiences discontinuities at integer values, including $J_{max}$.  Next, a graph of the imaginary part of the energy of an open system shows that there are discontinuities at integer values of particle number that are greater than or equal to $J_{max}$  (see Fig~\ref{fig:dis}).

\section{Conclusion}
As the particle number of a ground-state open system crosses an integer the exact complex Kohn-Sham exchange-correlation potential of DFRT extended to fractional particle numbers experiences a discontinuous jump.  At integers below the maximum number of bound electrons this jump is purely real and positive compensating for a purely real shift in the chemical potential.  At the specific integer that corresponds to the maximum number of bound electrons the jump has a non-zero imaginary component.  The real part of the jump is positive compensating again for the real shift in the chemical potential, and the imaginary part of the shift is negative due to the existence of a long-lived metastable state.

Since one can recover ground-state Density Functional Theory from DFRT by removing the complex transformation, we postulate that the exact exchange-correlation potential of DFT extended to fractional particle numbers should display a discontinuous jump at $J_{max}$, the maximum number of electrons the system can bind.  Due to the finite lifetime of the $(J_{max}+1)-$metastable state, the magnitude of this jump is larger than what could be expected by setting $A=0$.

\begin{acknowledgments}
The authors are grateful for valuable discussions with Mart\'in Mosquera.  Acknowledgment is made to the Donors of the American Chemical Society Petroleum Research Fund for support of this research under grant No.PRF\# 49599-DNI6.
\end{acknowledgments}

\bibliography{pplb_extension}

\end{document}
%